\documentclass[12pt,a4paper,titlepage,final]{article}
\pdfoutput=1
\usepackage[koi8-r]{inputenc}
\usepackage[russian]{babel}
\usepackage{verbatim}
\usepackage{amsfonts,amssymb,textcomp,amsmath}
\usepackage{graphicx}
\usepackage[sectionbib,square]{natbib}

\advance \textheight by \topskip
\begin{document}
\bibliographystyle{plainnat}

\vfill

\thispagestyle{empty}

\title{Поиск примеси стерильных нейтрино\\
при регистрации распадов трития\\
в пропорциональном счетчике:\\
новые возможности
}

\author{Д.Н.~Абдурашитов, А.И.~Берлев, Н.А.~Лиховид,\\А.В.~Лохов, И.И.~Ткачев, В.Э.~Янц}
\date{\it Институт ядерных исследований\\ \it Российской академии наук\\
\it Москва 117312, Россия}

\maketitle

\begin{abstract}
В настоящей работе предлагается эксперимент, направленный на поиск примеси стерильных нейтрино с массами в диапазоне 1--8~кэВ путем регистрации электронов от распада трития в пропорциональном счетчике. Примесь может быть обнаружена по специфическому искажению энергетического спектра электронов. Для указанных масс искажение распространяется на весь спектр, что позволяет использовать детекторы с относительно невысоким энергетическим разрешением. Новизна подхода заключается, с одной стороны, в использовании газового пропорционального счетчика с цельнопаяным кварцевым корпусом, позволяющим снимать токовый сигнал напрямую с анодной нити и обеспечивающим высокую стабильность при длительных измерениях. С другой стороны, применение современных цифровых методов регистрации позволит проводить измерения в условиях высокой загрузки ---  вплоть до 10$^6$~Гц. В результате энергетический спектр электронов от распадов трития, содержащий 10$^{12}$ отсчетов, можно будет набрать примерно за месяц. Такая статистика позволит поставить верхний предел на долю примеси стерильных нейтрино порядка \makebox[6mm][l]{10$^{-3}$}\ldots10$^{-5}$ на уровне достоверности 3$\sigma$ для $m_{s}$ в диапазоне 1--8~кэВ, что на 1\ldots2 порядка лучше современных значений.\\
\\
(Figure captions and some text are in English)
\end{abstract}

\section{Введение}
\label{sec:intro}
Ненулевая масса нейтрино и существование темной материи, природа которой неизвестна, свидетельствуют о том, что Стандартная модель физики частиц не полна. К числу наиболее естественных кандидатов на роль темной материи относятся стерильные нейтрино. Эти частицы появляются в большинстве расширений Стандартной модели на случай массивных нейтрино. Стерильные (правые) нейтрино это нейтральные лептоны. Являясь SU(2) синглетами они не участвуют в слабых взаимодействиях. Однако, за исключением специальных вырожденных случаев, они смешиваются с активными (левыми) нейтрино и приводят к появлению дополнительных массовых состояний в спектре. С экспериментальной точки зрения, в пользу существования стерильных нейтрино говорят результаты некоторых осцилляционных экспериментов с короткой базой (LSND и MiniBooNE), реакторная аномалия и результаты калибровки радиохимических экспериментов по измерению потоков солнечных нейтрино, см. обзор \citep{Abazajian:2012ys}. Эти аномалии можно описать в предположении одного или двух легких стерильных нейтрино с массой порядка электронвольт. Гипотеза о существовании дополнителнительных легких нейтрино не противоречит современным космологическим данным по измерениям анизотропии реликтового излучения \citep{Hinshaw:2012aka} и по распространенности легких элементов в свете теории первичного нуклеосинтеза \citep{Izotov:2010ca}. Более того,  эта гипотеза может снять \citep{Ade:2013lmv,Wyman:2013lza} существующие на сегодняшний день противоречия в наблюдениях Вселенной на малых и больших красных смещениях, а именно: между данными эксперимента Планк,  локальными измерениями постоянной Хаббла и имеющейся статистикой больших скоплений галактик. С точки же зрения объяснения темной материи, наиболее интересным является диапазон масс стерильных нейтрино в области 1--10~кэВ \citep{Dodelson:1993je,Boyarsky:2009ix}.

По аналогии с тремя поколениями фермионов Стандартной модели, можно ожидать, что стерильные правые нейтрино также образуют поколения. Обнаружение одного поколения стерильных нейтрино придаст уверенности в существовании остальных поколений. Если одно из них образует темную материю, соотношение между массой и углом смешивания такого нейтрино существенно ограниченo астрофизическими наблюдениями, такими как поиск гамма-линии от распада стерильных нейтрино в галактических гало \citep{Boyarsky:2006fg}, см. обзор \citep{Boyarsky:2009ix}. Эти ограничения не распространяются на остальные поколения. В связи с этим прямые лабораторные поиски стерильных нейтрино представляют особенный интерес, как в области значений параметров соответствующих сектору темной материи, так и вне ее. Обнаружение стерильных нейтрино даст ответ сразу на ряд фундаментальных вопросов физики частиц (структура массовой матрицы нейтрино, характер расширения стандартной модели, несохранение лептонного числа), астрофизики и космологии (темная материя).

\section{Поиск примеси массивных нейтрино в ядерных $\beta$-распадах}
\label{sec:beta}
Начало целенаправленных поисков примеси массивных нейтрино в ядерных $\beta$-распадах в лабораторных условиях можно увязывать с оригинальной работой Симпсона \citep{Simpson1985}, в которой было заявлено о наблюдении изгиба в спектре электронов трития в области 1.5~кэВ. Измерения проводились с использованием Si(Li) ППД, в который тритий был внедрен путем имплантации. На основании этого наблюдения был сделан вывод о существовании нейтрино с массой 17.1~кэВ и вероятностью смешивания ~3\%. После этого Симпсон с соавторами опубликовал еще 2 работы: спектр электронов трития, имплантированного в сверхчистый германиевый ППД \citep{Hime1989} и спектр электронов $^{35}$S, снятый внешним Si(Li) ППД, без имплантации источника в детектор \citep{Simpson1989}. В приведенных работах оценка на примесь была снижена до $\sim$0.6\%, при этом подтверждалось присутствие тяжелых нейтрино с массой порядка 17~кэВ. Эти результаты привели к всплеску экспериментов, направленных на поиск таких нейтрино в распадах $^{14}$C, $^{63}$Ni, $^{177}$Lu и некоторых других элементов. В целом ряде работ были получены в основном отрицательные результаты, исключавшие выводы Симпсона. Наиболее существенное на текущий момент ограничение на примесь массивных нейтрино было получено на магнитном спектрометре в Университете Цюриха по измерениям спектра электронов $^{63}$Ni. По этим измерениям верхний предел на примесь 17-кэвных нейтрино был оценен в 0.05\% на 95\% уровне достоверности \citep{Ni-Holzshun1999}.

\begin{figure}[tb]
\begin{center}
\includegraphics[width=1.0\textwidth]{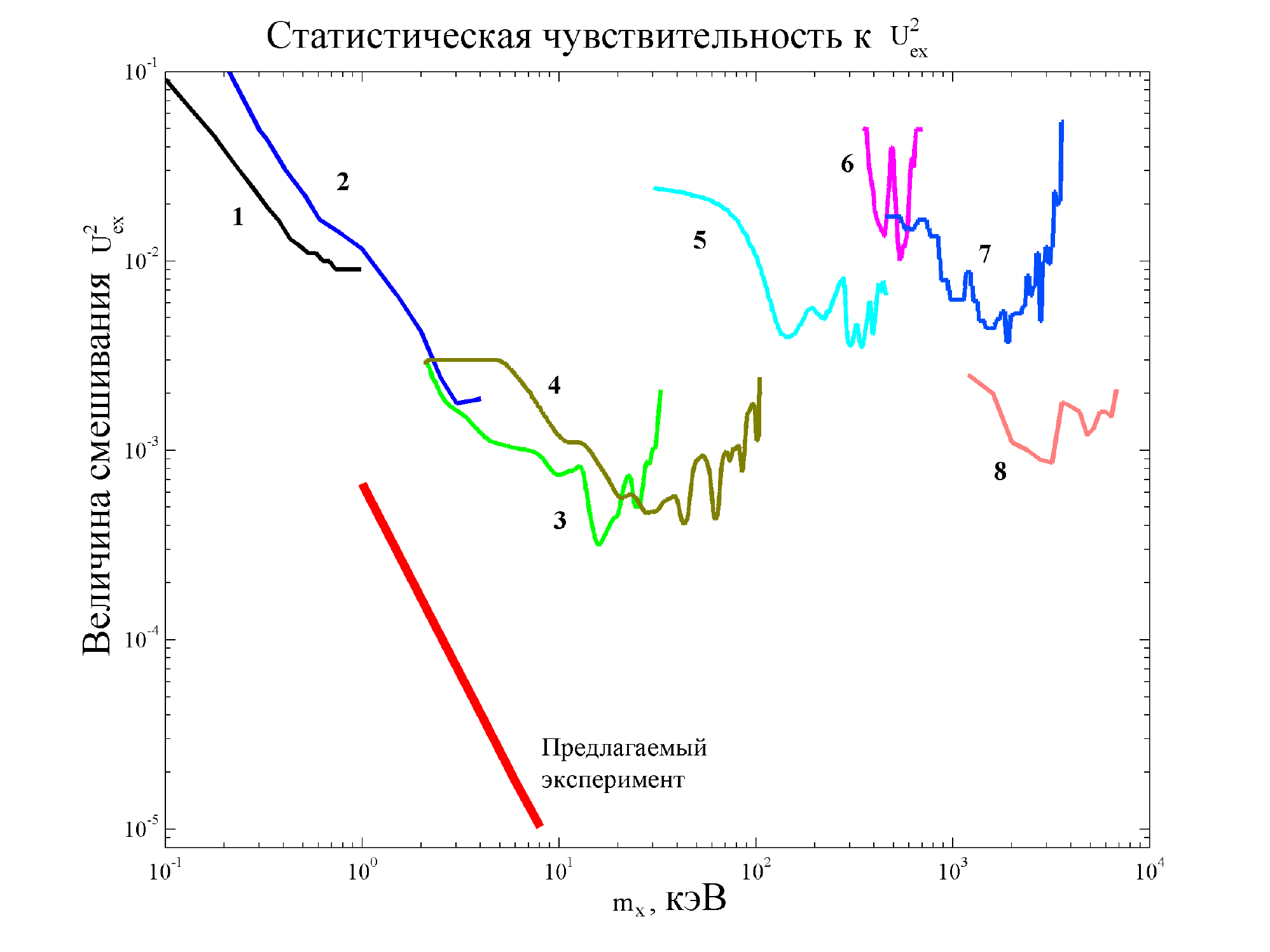}
\caption{Пределы на примесь стерильных нейтрино в ядерных распадах (нумерация пределов соответствует следующим экспериментам: 1 -- $^{187}$Re~\citep{Re-Galeazzi1978}; 2 -- $^3$H~\citep{T-Hiddeman1995}; 3 -- $^{63}$Ni~\citep{Ni-Holzshun1999}; 4 -- $^{35}$S~\citep{S-Holzschun2000}; 5 -- $^{64}$Cu~\citep{Cu-Shreck1983}; 6 -- $^{37}$Ar ядра отдачи~\citep{Ar-Hindi1998}; 7 -- $^{38m}$K~\citep{K-Trinczek2003}; 8 -- $^{20}$F~\citep{F-Deutsch1990}).}
\label{u2lim}
\end{center}
\end{figure}

Несмотря на то, что наличие примеси нейтрино с массой 17~кэВ на уровне 0.6\% было опровегнуто независимыми экспериментами, поиск массивных (стерильных) нейтрино в ядерных $\beta$-распадах в лабораторных условиях продолжился, хотя и не с такой интенсивностью. Современные ограничения на величину примеси отражены на Рис.~\ref{u2lim}, где представлены верхние пределы на значение параметра смешивания $U^2_{ex}$ для разных масс нейтрино m$_x$. Рассматривается простой случай смешивания массивных стерильных нейтрино $\nu_x$ с электронным флейворным состоянием активного нейтрино $\nu_e$. Рисунок подготовлен по опубликованным результатам лабораторных экспериментов последнего времени, в которых на основе детального анализа ядерных распадов проводился поиск примеси массивных стерильных нейтрино. Нумерация пределов соответствует следующим экспериментам: 1 -- $^{187}$Re~\citep{Re-Galeazzi1978}; 2 -- $^3$H~\citep{T-Hiddeman1995}; 3 -- $^{63}$Ni~\citep{Ni-Holzshun1999}; 4 -- $^{35}$S~\citep{S-Holzschun2000}; 5 -- $^{64}$Cu~\citep{Cu-Shreck1983}; 6 -- $^{37}$Ar ion recoils~\citep{Ar-Hindi1998}; 7 -- $^{38m}$K trap~\citep{K-Trinczek2003}; 8 -- $^{20}$F~\citep{F-Deutsch1990}. На этом рисунке не представлены пределы на примесь массивных нейтрино, полученные по тритию на магнитных спектрометрах в Троицке \citep{Trnumass2012} и в Майнце \citep{Mainz2012}. Эти эксперименты были нацелены в первую очередь на регистрацию $\beta$-спектра трития вблизи граничной энергии для точного измерения массы $\nu_e$, поэтому пределы на примесь получены для m$_x$ в диапазоне 1--100~эВ. Однако в ближайших планах этих групп (проект KATRIN как продолжение эксперимента в Майнце) стоят измерения спектра в более широком диапазоне, позволяющем поставить пределы для масс порядка нескольких кэВ.

На этом же рисунке показан верхний предел на параметр смешивания, ожидаемый в результате предлагаемого эксперимента (помечен текстом \flqq{Предлагаемый эксперимент}\frqq). Предел получен моделированием экспериментального спектра с обеспеченностью 10$^{12}$ событий; более подробно методика оценки описана ниже, в разделе \ref{sec:stat}. 

\section{Поиск примеси массивных нейтрино в $\beta$-распаде трития в пропорциональном счетчике}
\label{sec:counter}
С точки зрения методики обсуждаемого эксперимента наиболее интересны работы \citep{Bahran1992,Bahran1995}, в которых исследовался $\beta$-спектр трития в газовой пропорциональной камере. Работы выполнялись с целью проверки гипотезы 17-кэвных нейтрино, поэтому в основном на наличие изгиба исследовался участок спектра от 0.8 до 3.5~кэВ. Несмотря на это, некоторые методические особенности указанной работы имеют прямое отношение к предлагаемому эксперименту. В частности, была использована смесь Ar+10\%CH$_4$ при давлении 0.5~атм; максимальная скорость счета --- 2000~Гц. За 500~часов измерений статистическая обеспеченность по всему спектру составила примерно 10$^9$~событий, при этом в работе не сообщается о замеченной деградации газовой смеси. Далее, отмечено наличие стеночного эффекта, который в основном проявил себя в области малых (до $\sim$3~кэВ) энерговыделений. По этим измерениям верхний предел на примесь 17-кэвных нейтрино также был оценен в 0.2\% на 90\% уровне достоверности.

Следует отметить одну раннюю работу, также имеющую прямое отношение к предлагаемому эксперименту. Речь идет об измерениях массы электронного (анти)нейтрино по форме $\beta$-спектра трития вблизи граничной энергии. В работах \citep{Pontecorvo1948, Pontecorvo1949} был применен цилиндрический пропорциональный счетчик с газовой смесью на основе Xe+Ar+CH$_4$ в различных соотношениях. Активность трития в счетчике не превышала 30000~событий в минуту (500~Гц), о деградации смеси не сообщается. Также отмечается линейность шкалы и стабильность энергетического разрешения. Спектр трития исследовался в области выше 16~кэВ, ограничение на массу нейтрино составило 1~кэВ. С точки зрения методики важным моментом является также отмеченный пренебрежимо малый вклад стеночного эффекта в спектр электронов, начиная с 5~кэВ энерговыделения. И хотя численных оценок вклада стеночного эффекта не приведено, из этих работ можно сделать важный вывод о принципиальной возможности длительного набора спектра при высокой скорости счета в пропорциональном счетчике, так, что вклад от стеночного эффекта будет незначителен.

\section{Новый подход к измерениям спектра электронов от трития в пропорциональном счетчике}
\label{sec:new}
Как уже было отмечено выше, экспериментальный интерес представляет поиск специфического искажения энергетического спектра электронов в распадах трития, возникающего благодаря гипотетической примеси стерильных нейтрино с массами $m_{s}$ в диапазоне 1--10 ~кэВ. Как будет показано ниже, для указанных масс искажение распространяется на весь спектр, что позволяет использовать детекторы с относительно невысоким ($\sim$ 10-15\%) энергетическим разрешением. Для регистрации электронов будет использован цилиндрический газовый пропорциональный счетчик. Очевидным преимуществом газового счетчика в случае с тритием служит естественное совмещение источника с мишенью, что, в сочетании с низким порогом (вплоть до 0.2 кэВ), дает возможность регистрации практически всего спектра энерговыделений. Благодаря этому преимуществу, газовые детекторы неоднократно выбирались для измерения спектра трития \citep{ Bahran1992, Pontecorvo1948, Pontecorvo1949}. Предлагаемый для данной работы счетчик имеет специальную конструкцию --- выполняется из кварцевого стекла, с нанесением на внутреннюю поверхность тонкой углеродной пленки, служащей катодом \citep{Kuzminov1990, Yants1994}. Цельнопаяный кварцевый корпус обеспечивает, по сравнению с традиционными методами изготовления пропорциональных счетчиков, радикальное улучшение надежности и стабильности работы счетчика, а также возможность работы со счетчиком в широком диапазоне температур --- вплоть до 400~{\textdegree}C.

Предварительный выбор конструкции счетчика и смеси следующий. Счетчик представляет собой кварцевую трубку $\varnothing$10~мм и толщиной стенки 1~мм. Тонкий ($\sim$1~мкм) слой пиролитического графита (катод) наносится на внутреннюю поверхность трубки разложением изобутана; в качестве анода используется центральная вольфрамовая нить $\varnothing$12~мкм. При заполнении смесью Xe-CH$_4$ и при давлении $\sim$1~атм токовый импульсный сигнал, снимаемый в пропорциональном режиме напрямую с анодной нити, имеет длительность порядка 20~нс по основанию 10\%. Запись подробных (с шагом $\sim$1~нс) осциллограмм импульсов  обеспечит разрешение по времени порядка 3--5~нс и, как следствие, даст возможность работать с высокой скоростью счета.

Отличительной особенностью предлагаемого эксперимента служит возможность долговременных измерений в условиях высокой скорости поступления событий --- порядка \makebox[5mm][l]{10$^5$}\ldots10$^6$~Гц. Возможность длительного стабильного набора данных с высокой скоростью счета распадов $^3$H в пропорциональном счетчике обеспечивается, с одной стороны, подбором условий работы счетчика (оптимальная газовая смесь, не склонная к деградации, термостабилизация счетной аппаратуры и источников высокого напряжения и т.д.). С другой стороны, уровень развития современной цифровой обработки сигналов позволяет обеспечить накопление значительных объемов данных в режиме \flqq{online}\frqq. При этом не представляет технических затруднений запись формы каждого токового сигнала от счетчика, превысившего предустановленный порог, с дискретизацией порядка 1~нс.

Подобный режим набора реализован в ИЯИ РАН в исследованиях подбаръерного деления ядер на нейтронном спектрометре по времени замедления в свинце (СВЗ-100), а также в исследованиях резонансной структуры нейтронных сечений на времяпролетных спектрометрах РЭПС и РАДЭКС. Экспериментальная техника и метод измерения подробно описаны в работах \citep{Berlev2007,  Alexeev20101}. В качестве примера использования счетчиков с кварцевым корпусом можно привести эксперимент SAGE по измерению потока солнечных нейтрино \citep{sage2002}, в котором счетчики аналогичной конструкции стабильно работают в течение длительного времени.

Запись детальных осциллограмм импульсных сигналов с высоким разрешением по времени позволяет существенно снизить долю случайных наложений сигналов и осуществить набор значительной статистики в одном детекторе за разумное время. Так, при скорости 10$^6$~Гц за 1 месяц измерений будет набран спектр энерговыделений в счетчике, содержащий около 10$^{12}$~отсчетов в диапазоне 0.5--20~кэВ. Такая статистика позволит поставить верхний предел на долю примеси стерильных нейтрино порядка \makebox[6mm][l]{10$^{-3}$}\ldots10$^{-5}$ на уровне достоверности 3$\sigma$ для $m_{s}$ в диапазоне 1--8~кэВ. Указанный подход для поиска примеси стерильных нейтрино в распадах трития до настоящего времени не применялся. Кроме того, набор спектра электронов распада трития в одном детекторе со статистической обеспеченностью порядка 10$^{12}$ практически во всем диапазоне энергий будет уникальным измерением, не имеющим аналогов --- максимальное число событий в спектре трития, полученное когда-либо за одну серию измерений в пропорциональном счетчике, не превышает 10$^{9}$~\citep{Bahran1992}.

Методически описанный подход был уже реализован в работе \citep{Abd2006}. Однако главной целью этой работы была демонстрация возможностей метода записи осциллограмм импульсных сигналов на примере измерения активности $^{37}$Ar в пропорциональном счетчике в условиях интенсивной загрузки. Для решения этой задачи оказалось достаточно шага оцифровки 10~нс. Токовый импульсный сигнал с анода счетчика формировался предусилителем с характерным временем $\sim$20~нс. Длительность фронта сформированного сигнала составляла 50\ldots70~нс, спада --- до 1~мкс; временн\'ое разрешение составило 30~нс. В указанной работе не проводились оценки величин систематического искажения спектра $D$ и $C$ (см. раздел~\ref{sec:syst}), однако по косвенным данным эти параметры можно грубо оценить как $\sim$0.2\% и $\sim$1\% соответственно. В рамках методических исследований подобные измерения необходимо будет осуществить с упором на оценку именно этих параметров.

\section{Механизм формирования искажения спектра электронов при распаде трития}
\label{sec:analyt}
Тритий (Z = 1, A = 3) распадается по схеме
\[
^{3}\text{H} \:\rightarrow \:^{3}\text{He} + e^- + \bar{\nu}_{e},
 \label{eq: tritio}
\]
переходя в изотоп гелия $^{3}$He с периодом полураспада $t_{1/2} \simeq 12.3$~лет и энергией $ Q_{\beta}\simeq 18.6$~кэВ. Спин и четность ядер реакции не меняется ($ 1/2^{+} \rightarrow 1/2^{+} $), распад относится к сверхразрешенным фермиевским переходам и электрон испускается в виде $s$-волны. Для разрешенных переходов форма спектра испускаемых электронов определяется выражением
\[
\frac{dN}{dE} \sim F(Z,E) p E (E_0-E)^2.
 \label{eq:detritio1}
\]
Здесь $p,E$ --- импульс электрона и его энергия (в единицах $m_e c^2$), $E_0$ --- граничная энергия, а $F(Z,E)$ --- функция Ферми, учитывающая влияние кулоновского взаимодействия испускаемых электронов с ядром. Ненулевая масса нейтрино $m_{\nu}$ приводит к смещению граничной энергии спектра, которое описывается следующим образом:
\[
\frac{dN}{dE} \sim F(Z,E) p E (E_0-E) [(E_0-E)^2-(m_{\nu}c^2)^2]^{1/2}.
 \label{eq:detritionu}   
\]
В случае присутствия массивного стерильного нейтрино ${\nu}_{s}$ флейворное состояние ${\nu}_{e}$ (и, соответственно, $\bar{\nu}_{e} $) может быть записано как комбинация легкого активного (индекс $i$) и тяжелого стерильного массовых состояний:

\[
|\nu_{e}\rangle = \sum_{i} U_{ei} |\nu_{i}\rangle +  \sum_{s} U_{es} |\nu_{s}\rangle,
 \label{eq:mixing1}   
\]
где $U$ --- унитарная лептонная матрица смешивания. Без ущерба для общности можно аппроксимировать это выражение смешиванием только 2-х массовых состояний:

\[
|\nu_{e}\rangle = U_{el} |\nu_{l}\rangle + U_{es}|\nu_{s}\rangle.
\label{eq:mixing2}   
\]
Эффективная масса легкого активного нейтрино ($ m_{l} \lesssim $ eV) пренебрежимо мала по сравнению с массой искомого стерильного нейтрино порядка 1~кэВ. В этом случае можно записать спектр электронов распадов трития как \citep{Shrock1980}

\[
\frac{dN}{dE} =  (1-|U_{es}|^2) \frac{dN_{l}}{dE} + |U_{es}|^2 \frac{dN_{s}}{dE},
 \label{eq:spmix}   
\]
где

\[
\frac{dN_{l}}{dE} \sim F(Z=1,E) p E (E_0-E)^2
\]
 --- чистый спектр без примеси стерильного нейтрино и с нулевой массой активного нейтрино, а

\[
\frac{dN_{s}}{dE} \sim F(Z=1,E) p E (E_0-E) [(E_0-E)^2-(m_s c^2)^2]^{1/2}
\]
--- чистый примесный спектр со стерильным нейтрино. Здесь и в дальнейшем  мы пренебрегаем спектром конечных состояний молекулярного иона $^{3}$He.

С целью упрощения рассуждений рассмотрим только кинематическую составляющую формы спектра электронов. Для этого избавимся от функции Ферми в аналитических выражениях: чистый спектр без примеси стерильного нейтрино обозначим как $ S_{l} \equiv \frac{dN_{l}}{dE} / F(Z=1,E)$, чистый примесный спектр со стерильным нейтрино обозначим как $ S_{s} \equiv \frac{dN_{s}}{dE} / F(Z=1,E)$, а смешанный спектр --- $ S \equiv \frac{dN}{dE} / F(Z=1,E)$. В этом случае возможное искажение спектра испускаемых электронов может выглядеть следующим образом.

\begin{figure}[tb]
\begin{center}
\includegraphics[width=1.0\textwidth]{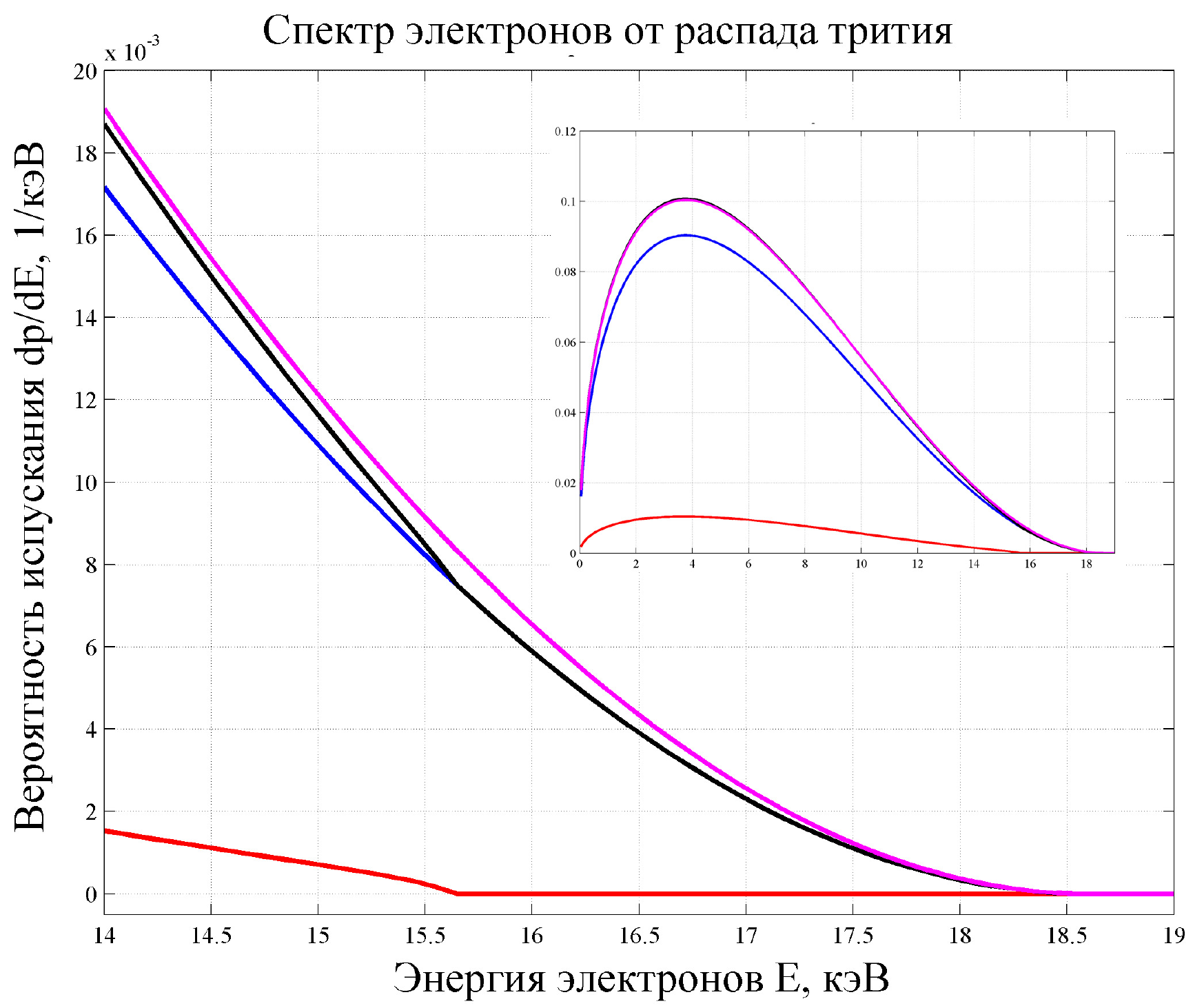}
\caption{Спектр электронов трития с примесью стерильных нейтрино на участке 14--19~кэВ: $U^2$=0.1, $m_s$=3~кэВ; лиловый цвет -- чистый спектр $S_l$ без примеси $\nu_s$, синий цвет -- этот же спектр с коэффициентом $1-0.1=0.9$, красный цвет -- чистый примесный спектр $S_s$ с коэффициентом 0.1, смешанный спектр $S =0.9 S_l + 0.1 S_s$ показан черным цветом. На вкладке представлен спектр электронов во всем диапазоне энергий.}
\label{stmix}
\end{center}
\end{figure}

На Рис.~\ref{stmix} показаны составляющие смешанного спектра для искусственного случая $U^2$ = 0.1, $m_s$ = 3~кэВ. Лиловым цветом представлен чистый спектр $S_l$ (без примеси $\nu_s$), синим цветом -- этот же спектр с коэффициентом $1-0.1=0.9$. Красный цвет соответствует чистому примесному спектру $S_s$ с коэффициентом 0.1; смешанный спектр $S =0.9 S_l + 0.1 S_s$ показан черной линией. Спектры нормированы на 1, так, что представление по смыслу соответствует плотности вероятности испускания электрона $df$ с энергией в бине шириной $dE$=1~кэВ. Следует отметить, что даже для такого сильного смешивания спектр $S$ практически не отличается от $S_l$; это хорошо видно на вкладке, где представлены спектры во всем диапазоне энергий -- черная линия сливается с лиловой. Основное искажение спектра наблюдается на участке, отстоящем от граничной энергии $E_0$ на величину $m_s$ = 3~кэВ -- изгиб черной линии в области 15.6~кэВ.
\label{Stat}

\begin{figure}[tb]
\begin{center}
\includegraphics[width=1.0\textwidth]{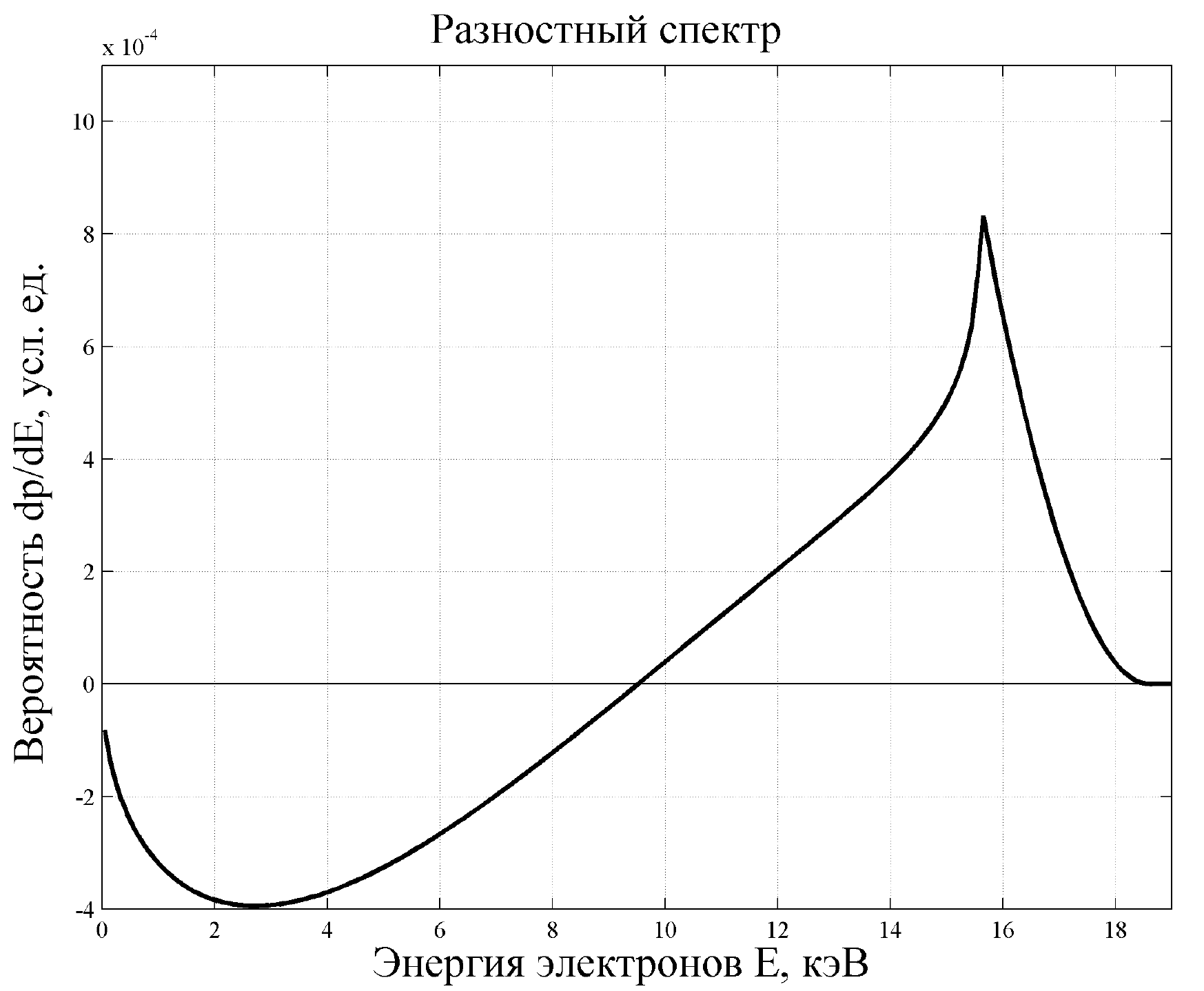}
\caption{Разность $S_l - S$ для искусственного случая $U^2$ = 0.1, $m_s$ = 3~кэВ.}
\label{tdiff}
\end{center}
\end{figure}

На Рис.~\ref{tdiff} показана разность функций плотностей вероятности испускания $S_l - S$, наглядно характеризующая особенности изгиба -- острый пик в точке изгиба на фоне искажения формы по всему спектру. 

\section{Чувствительность метода: статистическая неопределенность}
\label{sec:stat}
Описанное выше поведение дает возможность использовать для  оценки  величины искажения интегралы по разностному спектру $S_l-S$. Проблема состоит в том, что измеряемый спектр $S_{meas}$ является суперпозицией не содержащего эффекта спектра $S_l$ и собственно эффекта $S_s$, при этом возможность измерить \flqq{эталонный}\frqq\ (беспримесный) спектр $S_l$ принципиально отсутствует. Однако, эта трудность, типичная для широкого круга физических задач, в данном случае смягчается различным влиянием эффекта на разные области спектра. В частности, для оценки параметров беспримесного спектра выглядит целесообразным использовать интервал, соответствующий отрицательной величине разности (большая статистика, меньшее искажение формы спектра по отношению к эталонному). Тем самым \flqq{отрицательный}\frqq\ интервал становится эталонным, соответственно \flqq{положительный}\frqq\ интервал будет контрольным. 
Нас будет интересовать разница в предполагаемой (на основании фитирования эталонного интервала беспримесным спектром) и фактической скорости счета на контрольном интервале. Предполагаемый эффект будет складываться как из собственно искажения спектра, так и из смещения оценки вследствие влияния примеси стерильного нейтрино  на спектр в эталонном интервале. В данном случае эти вклады будут иметь один знак, то есть усиливать эффект. Фактически речь будет идти о проверке гипотезы отсутствия искажений в измеренном спектре.

Очевидно, что граница между эталонным и контрольным интервалами будет зависеть от массы стерильного нейтрино, что в принципе (при наличии эффекта), позволит также оценить ее величину. Поэтому полная процедура обработки в этом подходе должна включать максимизацию эффекта путем подбора оптимальных границ эталонного и контрольного интервалов. Однако, в связи с предварительным характером оценок, при обсчете результатов моделирования эталонный и контрольный интервалы были фиксированными (для упрощения расчетов). Таким образом, приводимые ниже оценки чувствительности являются скорее заниженными.

Эталонным интервалом был выбран участок от 4 до 10~кэВ, содержащий $\sim60\%$ от всего спектра. Нижняя граница участка установлена достаточно высоко с целью ограничить влияние неопределенности функции Ферми $ F(Z=1,E)$. Выбор интервала фитирования вполне произволен и имеет предварительный характер. Получив оценку беспримесного спектра $S_{l}^{fit}$, можно анализировать разность $S_{l}^{fit} - S_{meas}$ на контрольном интервале 10-18~кэВ. Выбор указанных границ интервалов ограничивает диапазон поиска стерильных нейтрино массами 1-8~кэВ.

Высокая скорость счета позволит за разумное время набрать спектр энерговыделений со статистической обеспеченностью порядка 10$^{12}$ событий. Оценим, какой предел на вероятность примеси стерильных нейтрино можно будет поставить на такой статистике. Как уже отмечалось, мерой искажения смешанного спектра можно в первом приближении выбрать  интеграл $\mathcal{J}$ разностного спектра на контрольном интервале 10-18~кэВ: $\mathcal{J} = \int_{10}^{18} (S_{l}^{fit} - S_{meas}) dE$. Из выражения для $\frac{dN}{dE}$ ясно, что $\mathcal{J}$ и $U^2$ связаны линейно. Численные значения $\mathcal{J}$, выраженные в долях нормированного спектра для различных масс стерильного нейтрино, представлены в Табл.~\ref{delta}.

\begin{table}[tbp]
\begin{center}
\begin{tabular}{|c|ccccc|}
\hline
$U^2_{es}$ &\multicolumn{5}{c|}{Масса $\nu_s$, кэВ}\\
\cline{2-6}
& 1 & 2 & 4 & 6 & 8 \\
\hline
10$^{-3}$ & 2.7$\cdot10^{-6}$ & 1.1$\cdot10^{-5}$ & 4.4$\cdot10^{-5}$ & 1.0$\cdot10^{-4}$  & 1.7$\cdot10^{-4}$ \\
10$^{-4}$ & 2.7$\cdot10^{-7}$ & 1.1$\cdot10^{-6}$ & 4.4$\cdot10^{-6}$ & 1.0$\cdot10^{-5}$  & 1.7$\cdot10^{-5}$ \\
10$^{-5}$ & 2.7$\cdot10^{-8}$ & 1.1$\cdot10^{-7}$ & 4.4$\cdot10^{-7}$ & 1.0$\cdot10^{-6}$  & 1.7$\cdot10^{-6}$ \\
\hline
\end{tabular}
\caption{Значения $\mathcal{J}$, выраженные в долях нормированного спектра для различных масс нейтрино и нескольких значений $U^2_{es}$}
\label{delta}
\end{center}
\end{table}

В случае отсутствия сигнала для каждого значения массы $m_s$ находится наименьшее значение $U^2$, при котором $\mathcal{J}$ все еще превышает статистическую неопределенность на уровне 3$\sigma$. Для статистики 10$^{12}$ этот уровень на интервале 10-18~кэВ составляет $3 \cdot \sqrt{0.35 \cdot 10^{12}} \sim 1.8 \cdot 10^6$ отсчетов, что соответствует доле спектра $\sim 1.8 \cdot 10^{-6}$ \label{statlim}. Из Табл.~\ref{delta} видно, что для $m_s$ = 1~кэВ только при смешивании порядка $U^2 = 10^{-3}$ значение $\mathcal{J}$ оказывается выше этого уровня. Другими словами, на статистике 10$^{12}$ событий верхний предел на $U^2$ (или статистическая чувствительность эксперимента) для $m_s$ = 1~кэВ составит $10^{-3}$. Для массы $m_s$ = 8~кэВ верхний предел на смешивание составит $\sim10^{-5}$.

Необходимо заметить, что изложенный подход к оценке пределов на данной статистике достаточно груб и носит предварительный характер, как и в случае с выбором эталонного интервала энергий для фитирования беспримесного спектра. Детальное моделирование покажет, можно ли повысить чувствительность, анализируя разностный спектр на предмет специфических особенностей, таких как, например, наличие острого пика (см. Рис.~\ref{tdiff}).

Вообще говоря, для поиска примеси стерильных нейтрино и оценок чувствительности можно также использовать прямое фитирование спектра с добавлением двух дополнительных параметров - массы тяжелого нейтрино и квадрата синуса угла смешивания. Однако такой способ имеет сильную зависимость от знания теоретического спектра. Более удобным может оказаться применение специальных статистических критериев, построенных, например, с помощью метода квазиоптимальных весов \citep{FTkachev2006}. Можно показать, что такие критерии достаточно чувствительны к наличию вклада стерильных нейтрино в спектре, но не требуют точного знания теоретического спектра. Такие критерии были построены, например, в \citep{Lokhov2012} для поиска аномального вклада типа ступеньки в спектре электронов $\beta$-распада трития в эксперименте \flqq{Троицк-$\nu$-масс}\frqq.

\section{Систематическое искажение амплитудного спектра пропорционального счетчика}
\label{sec:systmodel}
Факторы систематического воздействия на отклик цилиндрических газовых счетчиков хорошо известны. Основной вклад в отклик счетчика вносит разброс измеряемых амплитуд сигналов, обусловленный в первую очередь статистическим разбросом числа $n$ ион-электронных пар первичной ионизации. Исторически этот разброс характеризуется параметром $R$ (энергетическое разрешение), который по определению равен отношению ширины на $1/2$ высоты (full width at half maximum, f.w.h.m.) отклика счетчика на линию к энергии линии: $R = \frac{f.w.h.m.}{E}$. Полная ширина на полувысоте и дисперсия нормального распределения $\sigma$ связаны соотношением f.w.h.m.=$2.34\cdot\sigma$; оценкой дисперсии для больших чисел служит $\sqrt{n}$. В этом случае $R=2.34\cdot\frac{\sqrt{n}}{n}=\frac{2.34}{\sqrt{n}}$; в пропорциональном режиме, с газовым усилением порядка 10$^3$, соответственно $R\propto\frac{1}{\sqrt{E}}$, где $E$ --- энергия электронов.  Для инертных газов средняя энергия, затрачиваемая на образование иона составляет примерно 30~эВ. Электроны с энергией 5.9~кэВ, выбиваемые в рабочем газе счетчика рентгеновским источником $^{55}$Fe, часто применяемого для калибровки, образуют в среднем $\sim$200 электронов первичной ионизации с дисперсией $\sigma\sim\sqrt{200}\sim14$ электронов. Таким образом, отклик счетчика на электроны с энергией 5.9~кэВ в лучшем случае обладает разрешением $R_{5.9}=\frac{2.34}{14}\sim16\%$; для энергии 18~кэВ разрешение составит около 10\%.

Значительный вклад в систематику обусловлен формированием подложки в спектре за счет стеночного и краевого эффектов, и характеризуется долей $D$ событий,\label{degr} ушедших в подложку. В рабочей смеси с преобладанием ксенона при 1~атм размер облака первичной ионизации от электрона с энергией $\sim$10~кэВ имеет порядок $l\sim$0.1~мм. В этом случае в счетчике с характерным поперечным размером $L\sim$10~мм доля событий с деградацией энергии (стеночный эффект) составит $\frac{l}{L}\sim$1\%. Специальная конструкция счетчика (конусная форма торцов) и достаточная длина ($\sim$200~мм) позволяет довести долю краевого эффекта до величины  порядка $\sim$0.1\% (см., например, \citep{Abd1994}).

При высокой скорости счета начинают играть роль случайные совпадения (наложения) событий. Скорость счета совпавших событий $C$ прямо пропорциональна произведению скорости счета и длительности сигнала. Запись формы каждого сигнала обеспечивает разрешение по времени порядка 3{\ldots}5~нс, и при скорости счета $\sim$10$^6$~Гц доля неразличимых наложений  $C$, искажающих конечный амплитудный спектр, составит величину $\sim$0.1\%.

\begin{figure}[tb]
\begin{center}
\includegraphics[width=1.0\textwidth]{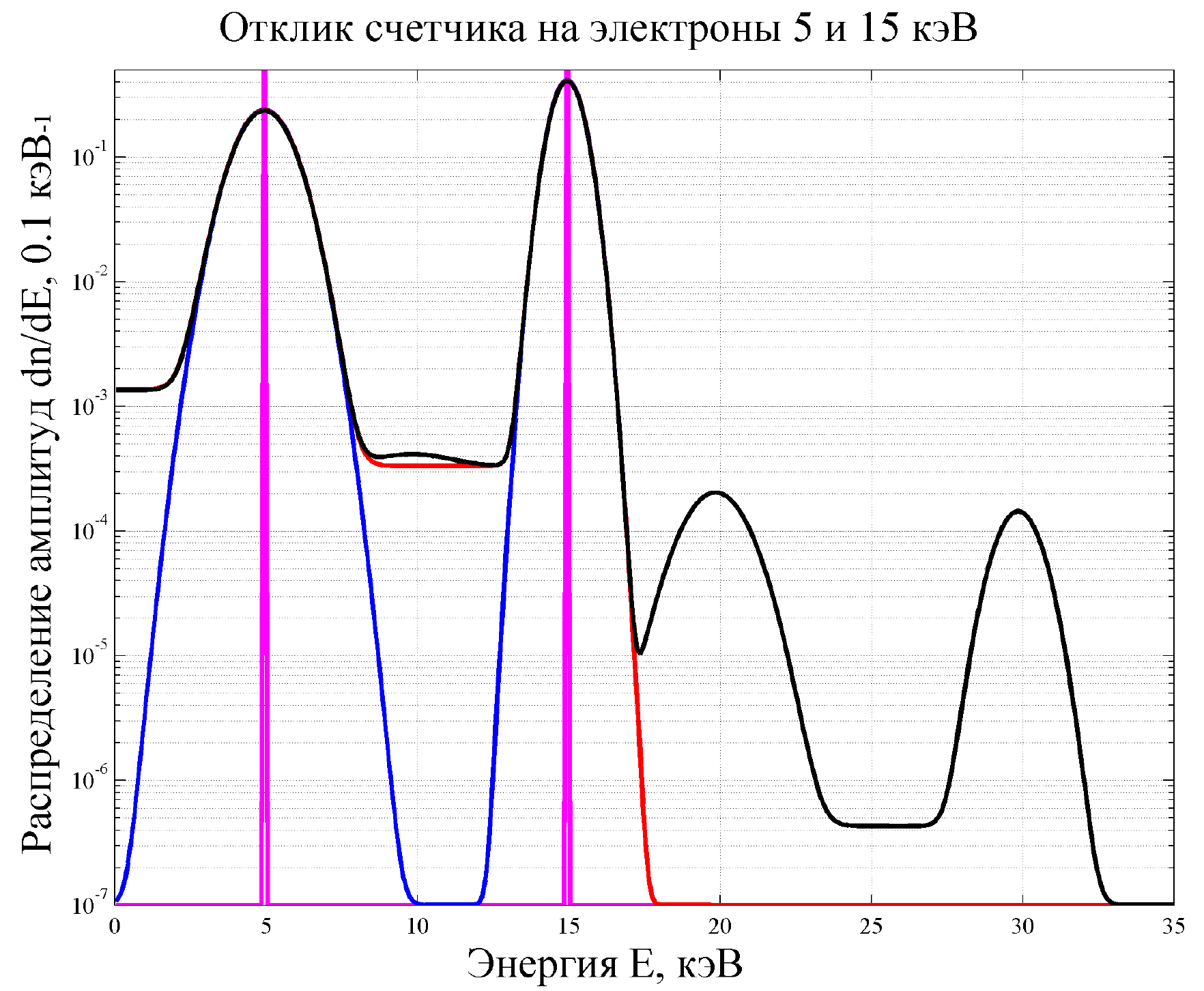}
\caption{Модельный отклик счетчика на электроны с энергией 5 и 15~кэВ (лиловый цвет), искаженный разрешением, деградацией (стеночный и краевой эффекты) и неразличимым наложением событий. Синий цвет -- $R_{5.9}=16\%$, красный цвет -- $D=1\%$, черный цвет -- $C=0.1\%$.}
\label{2line}
\end{center}
\end{figure}

\begin{figure}[tb]
\begin{center}
\includegraphics[width=1.0\textwidth]{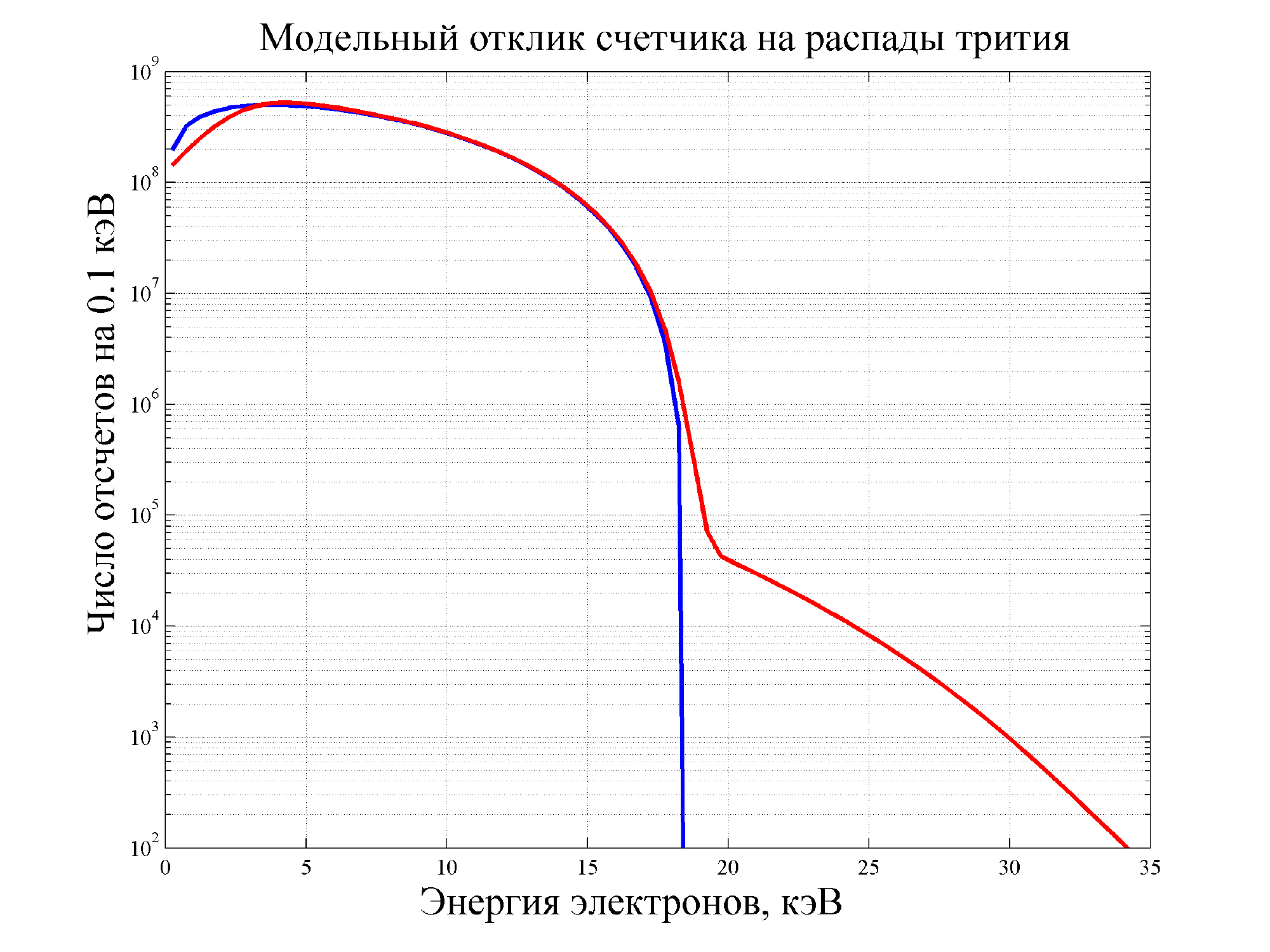}
\caption{Модельный спектр событий в счетчике в присутствии трития для статистики N=10$^{12}$. Синий цвет -- исходный спектр испускаемых электронов, красный цвет -- амплитудный спектр, искаженный разрешением ($R_{5.9}=16\%$), деградацией ($D=1\%$) и неразличимым наложением событий ($C=0.1\%$).}
\label{tritmod}
\end{center}
\end{figure}

Искажение отклика пропорционального счетчика, обусловленное разрешением, деградацией и неразличимым наложением событий, проиллюстрировано рисунком~\ref{2line}, на котором представлено модельное распределение амплитуд сигналов счетчика. Для наглядности показан отклик на электроны энергий 5 и 15~кэВ. Цветовое обозначение линий следующее: лиловый цвет -- исходные линии электронов, каждая интенсивностью 1; синий цвет -- размытие с разрешением $R_{5.9}=16\%$; красный цвет -- искажение отклика за счет краевого и стеночного эффектов ($D=1\%$); черный цвет соответствует неразличимым наложениям событий с коэффициентом $C=0.1\%$. Амплитудный спектр сигналов счетчика от распадов трития, искаженный таким откликом, представлен на Рис.~\ref{tritmod}.

\section{Чувствительность метода: cистематические факторы}
\label{sec:syst}
Чувствительность эксперимента, помимо статистической погрешности, зависит также и от величины систематической неопределенности. Количественную оценку воздействия систематических факторов искажения спектра, описанных выше, можно получить, проводя моделирование.

Экспериментальный спектр $S$ с определенной статистикой $N$ и при заданных значениях $m_s$ и $U^2$ случайным образом формируется, используя аналитические выражения для плотности вероятности испускания электронов (см. разд.~\ref{sec:analyt}). Затем полученный спектр искажается разрешением $R$, подложкой $D$ и наложениями $C$. Таким образом, смоделированный спектр $S$ является функцией указанных параметров:

\[
S=S(m_s, U^2, N, R, D, C).
\]
Следующим этапом проводится фитирование (подгонка) $S$ чистым беспримесным спектром $S_{l}^{fit}$, искаженным соответствующей систематикой:
\[
S_{l}^{fit}=S_{l}^{fit}(N,R,D,C).
\]
При этом параметры $D$ и $C$ фиксируются, а $N$ и $R$ оставляются свободными, т.е. в результате подгонки получаются конкретные значения $N^{fit}$ и $R^{fit}$, с которыми беспримесный спектр наиболее удовлетворительно укладывается в экспериментальный. Для подгонки использовалась стандартная функция нелинейного фита nlinfit() из пакета Matlab~7.0; фитирование проводилось на участке 4-12~кэВ. В конечном итоге из разности $S_{l}^{fit} - S$ получается оценка сигнала $\mathcal{J}^{fit}$ как интеграл разностного спектра на участке 12-18~кэВ.

С точки зрения оценки систематической неопределенности интерес представляет разброс значений $\mathcal{J}$ при вариациях фиксированных параметров $D$ и $C$. Разброс возникает в том случае, когда модельный спектр $S$ получается при одних значениях $D$ и $C$, а фитирование проводится с другими значениями $D\pm\delta{D}$ и $C\pm\delta{C}$. Такая ситуация моделирует неточное знание $D$ и $C$ в реальном эксперименте и, следовательно, приводит к оценке систематической неопределенности.

\begin{figure}[tb]
\begin{center}
\includegraphics[width=1.0\textwidth]{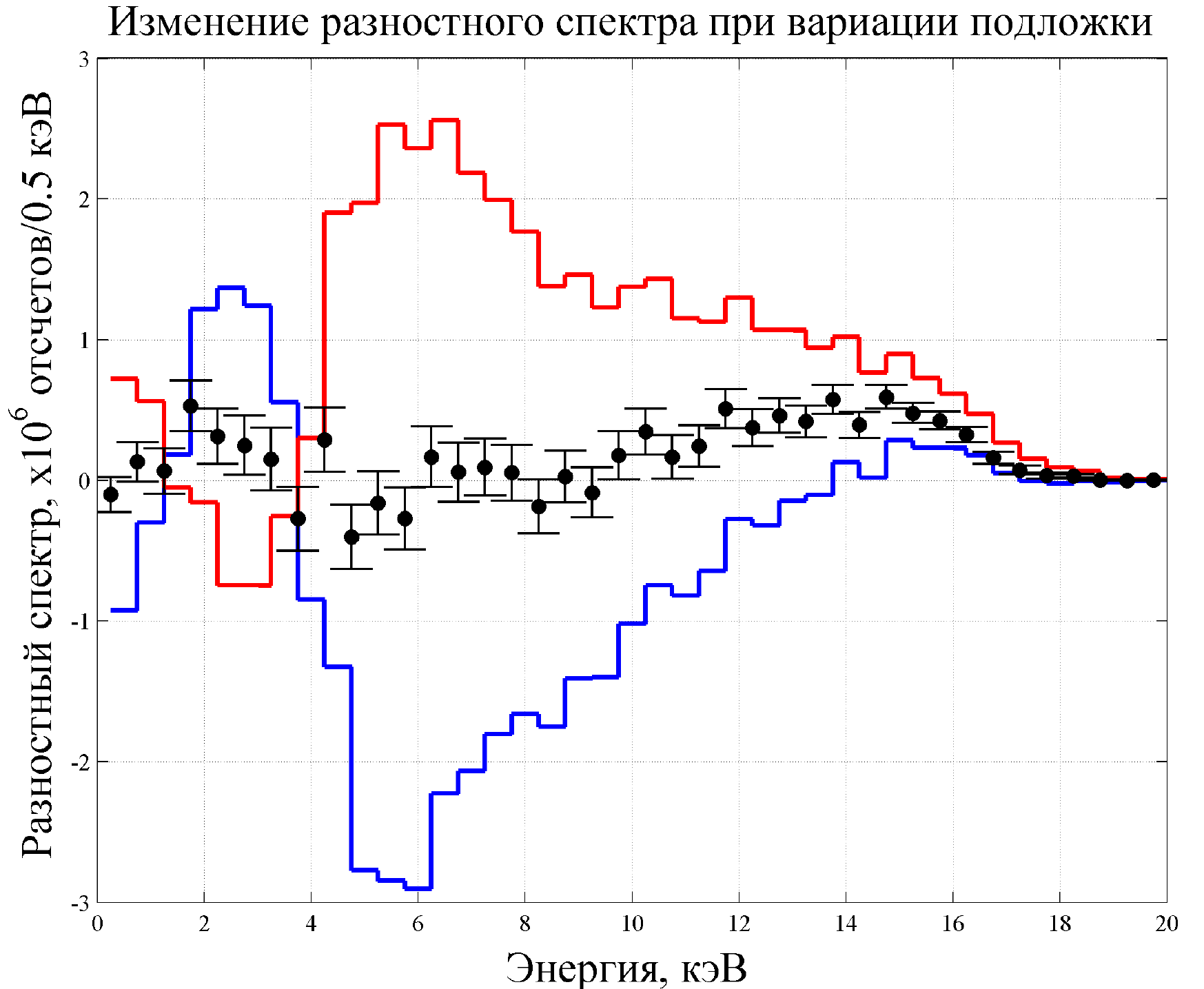}
\caption{Разностный спектр $S_{l}^{fit} - S$ с вариациями подложки при центральных значениях $D_0$=1\% и $C_0$=0.1\% ($U^2$ = 10$^{-4}$,  $m_s$ = 4~кэВ, N=10$^{12}$). Точки с ошибками -- фит с $D$=$D_0$, $C$=$C_0$; синяя линия -- $D$=$D_0\cdot[1-0.01]$; красная линия -- $D$=$D_0\cdot[1+0.01]$.}
\label{systD}
\end{center}
\end{figure}

\begin{figure}[tb]
\begin{center}
\includegraphics[width=1.0\textwidth]{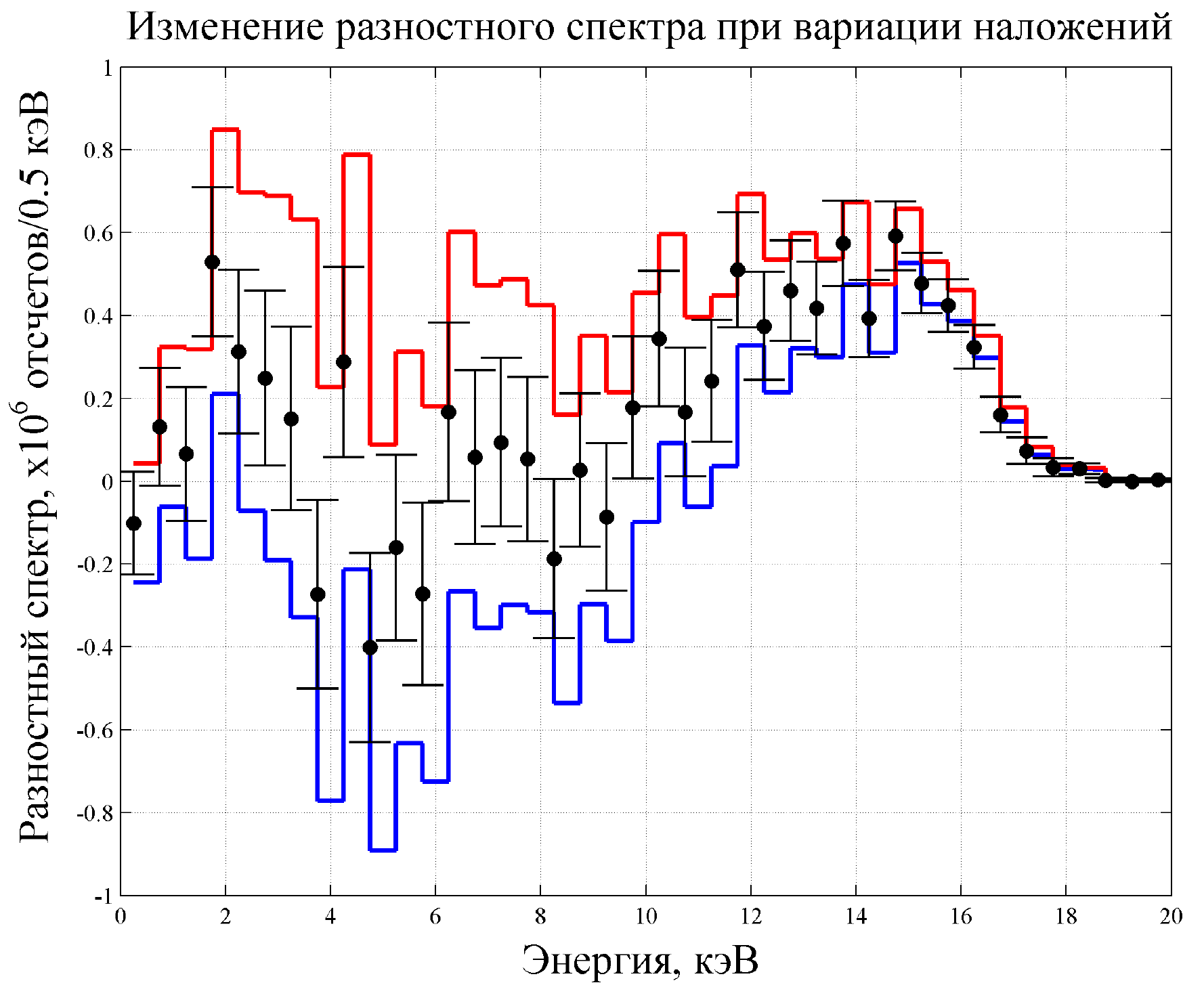}
\caption{Разностный спектр $S_{l}^{fit} - S$ с вариациями наложений при центральных значениях $D_0$=1\% и $C_0$=0.1\% ($U^2$ = 10$^{-4}$,  $m_s$ = 4~кэВ, N=10$^{12}$). Точки с ошибками -- фит с $D$=$D_0$, $C$=$C_0$; синяя линия -- $C$=$C_0\cdot[1-0.01]$; красная линия -- $C$=$C_0\cdot[1+0.01]$.}
\label{systC}
\end{center}
\end{figure}

В качестве иллюстрации изменения модельного отклика счетчика при вариациях подложки и наложений приведем результаты моделирования разностных спектров для случая $U^2$ = 10$^{-4}$ и $m_s$ = 4~кэВ; модельная статистика N=10$^{12}$. На Рис.~\ref{systD} представлен разностный спектр $S_{l}^{fit} - S$ с одинаковыми $D$ и $C$ при центральных значениях $D_0$=10$^{-2}$ и $C_0$=10$^{-3}$ (показан точками с ошибками). Здесь же показаны разности в случае, когда фит проводится при других значениях подложки: $D$=$D_0\cdot[1-0.01]$ (синяя линия) и $D$=$D_0\cdot[1+0.01]$ (красная линия). Аналогично этому, на Рис.~\ref{systC} представлен тот же модельный разностный спектр вместе с фитом при других значениях наложений: $C$=$C_0\cdot[1-0.01]$ (синяя линия) и $C$=$C_0\cdot[1+0.01]$ (красная линия). Из приведенных рисунков отчетливо видно, что даже при небольшой, порядка 1\%, неточности в определении $D$ и $C$ сигнал может искажаться значительно, особенно в случае подложки.

\begin{figure}[tb]
\begin{center}
\includegraphics[width=1.0\textwidth]{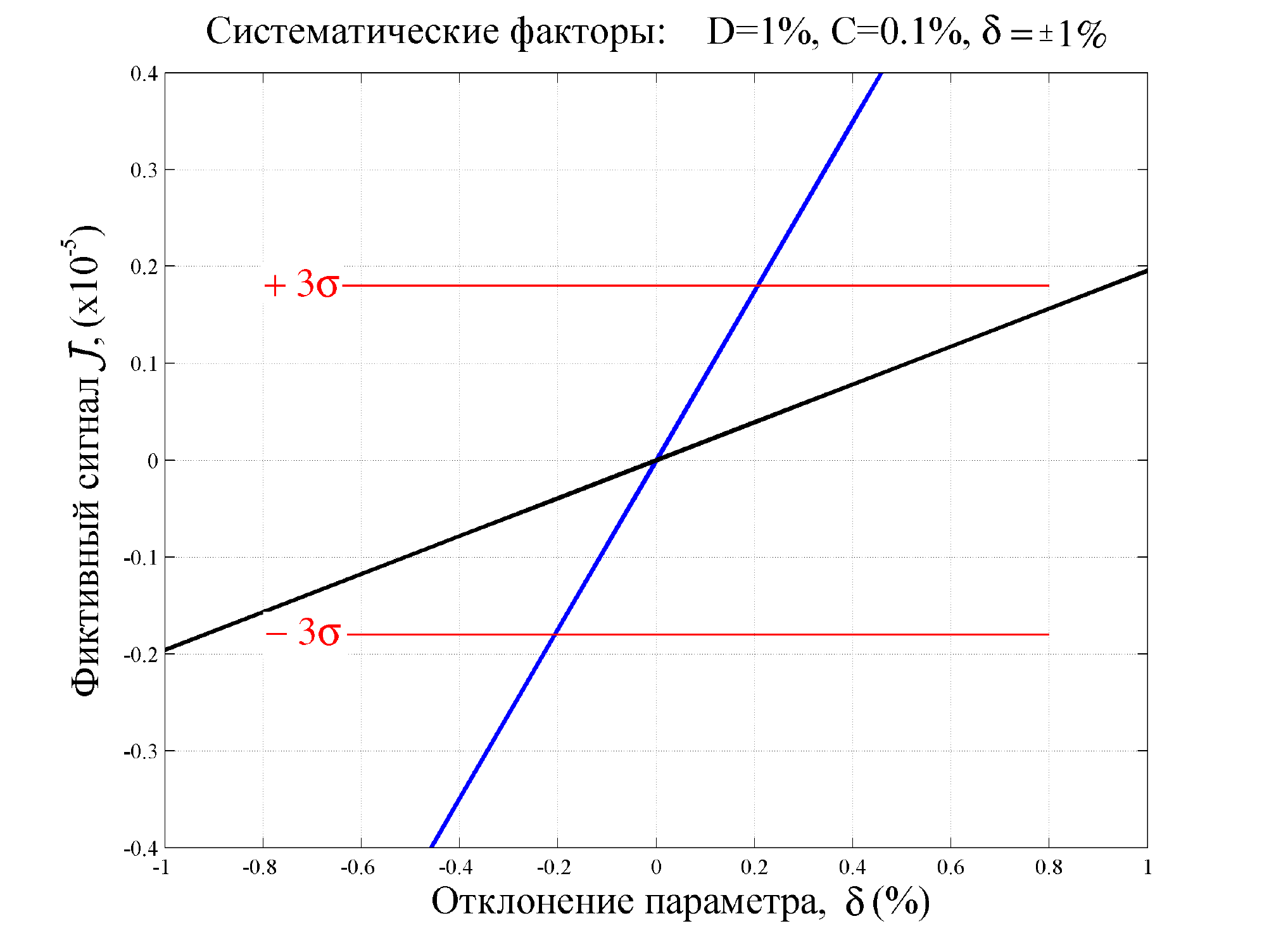}
\caption{Значения фиктивного сигнала $\mathcal{J}$ при нулевом смешивании в зависимости от изменения подложки $D$ (синяя линия) и наложений $C$ (черная линия). Красные линии -- границы статистической неопределенности нулевого сигнала для N=10$^{12}$, соответствущей 1.8$\cdot10^6$ отсчетов.}
\label{systDC}
\end{center}
\end{figure}

Оценка систематической неопределенности, связанной с неточным знанием подложки и наложений, была проведена путем моделирования в отсутствие сигнала стерильных нейтрино в смешанном спектре (случай $U^2$ = 0). Параметры $D$ и $C$ варьировались в пределах $\pm$1\% относительно центральных значений $D_0$=10$^{-2}$ и $C_0$=10$^{-3}$ на статистике N=10$^{12}$. Для каждого моделирования фиксировалось значение фиктивного сигнала $\mathcal{J}$, порожденного систематическим воздействием. Результаты сведены в Рис.~\ref{systDC}, на котором представлены значения сигнала в зависимости от изменения $D$ (синяя линия) и от изменения $C$ (черная линия). Две красные линии отмечают границу статистической неопределенности нулевого сигнала на уровне 3$\sigma$, которая для N=10$^{12}$ соответствует 1.8$\cdot10^6$ отсчетов (см. разд.~\ref{sec:stat}).

Из этого рисунка можно сделать вывод, важный с точки зрения планирования эксперимента. Неточность знания подложки (параметр $D$) воздействует на конечный результат значительно сильнее по сравнению с неточностью наложений $C$. В таком случае, чтобы удержать систематическую неопределенность в пределах статистической, необходимы калибровки, обеспечивающие измерения подложки с точностью $\sim$0.2\% и наложений с точностью $\sim$1\%.

\section{Обсуждение}
\label{sec:discuss}
Минимизация традиционных факторов систематической погрешности предполагает подбор конструкции счетчика и параметров рабочей смеси, обеспечивающих стабильность его работы в условиях высокой загрузки. Необходимо будет изготовить кварцевые счетчики с углеродным катодным напылением различных диаметров и длины; для различных смесей инертных газов (Ar, Xe и др.) с гасящими добавками CH$_4$, CO$_2$ и др. исследовать оптимальные состав и давление, обеспечивающие быстродействие и долговременную стабильность в изготовленных счетчиках. На основе измерений можно будет выбрать наиболее подходящий счетчик и состав смеси для проведения измерений с $^3$H. Детальные расчеты (численное моделирование) будут необходимы для уточнения вклада различных источников систематической неопределенности. Для калибровочных измерений можно использовать источники внутренние ($^{37}$Ar, 2.8~кэВ) и внешние --- $^{55}$Fe, 5.9~кэВ и $^{109}$Cd, 22~кэВ. Следует отметить, что измерения спектра трития предполагается проводить при максимальной загрузке, поэтому калибровать счетчик с рабочим тритиевым заполнением не представляется возможным. По этой причине линейность рабочей шкалы, величину энергетического разрешения и стабильности счетчика оптимально оценивать при помощи калибровочных источников до и после измерений, с аналогичной смесью.

Как показывают предварительные оценки, при средних значениях $D{\sim}10^{-2}$ и $C{\sim}10^{-3}$ с точностью $\pm0.1\%$ и $\pm1\%$ соответственно вклад систематической неопределенности в оценку предела на примесь стерильных нейтрино оказывается сравним с вкладом статистической неопределенности на уровне 3$\sigma$ при общем числе событий в спектре порядка 10$^{12}$. Такая точность измерений подложки может быть достигнута при статистике калибровочных спектров не менее 10$^8$ событий, что не представляет проблем. Необходимо, однако, отметить, что простая модель прямоугольной подложки, заложенная в оценку систематики, скорее всего, не полностью описывает подложку. Прямые калибровочные измерения с большой статистикой позволят уточнить истинную форму подложки.

Специфическим для измерений с тритием фактором систематической неопределенности будет вклад распадов атомов, поглощенных стенками счетчика. По имеющимся данным \citep{Bahran1992} при использовании не содержащих водород материалов и соответствующей методике измерений, вклад этого фактора достаточно мал, и может быть проконтролирован и корректно учтен. Кроме того, ряд специалистов по взаимодействию трития с материалами (напр., \citep{Kaz2013}), сообщают, что водород и тритий в основном сформирует мономолекулярный слой на плоской поверхности пиролитического графита, нанесенного на внутреннюю поверхность кварцевой трубки. Это одно из преимуществ применения кварцевого счетчика, в отличие от металлического катода, в котором в первую очередь ожидается диффузия водорода с поверхности в глубину металла.

\section{Заключение}
В заключение авторы выражают признательность Н.Титову, В.Пантуеву, В.Матушко и А.Нозику за плодотворное обсуждение и конструктивную критику предлагаемого эксперимента, а также С.Гирину за внимательное прочтение и правку текста.

\newpage
\section*{Figure captions and some text in English}
\label{sec:eng}

This is a proposal of very simple measurement so mostly it is described in the abstract. In addition let us comment figures.

Fig.1 - red line is an upper limit on mixing of the experiment due to statistics of 10$^{12}$ events.

Fig.2 - the normalized spectrum of tritium electrons in case of m$_s$=3 keV and U$^2$=0.1 is presented; magenta line is pure spectrum without sterile neutrino S$_l$; blue line is the same one but reduced by 0.9 factor; red line is pure sterile neutrino spectrum S$_s$ reduced by 0.1; black line is mixed one S=0.9xS$_l$ + 0.1xS$_s$.

Fig.3 - the difference (S$_l$ - S) is shown; we derive a J as the (S$_l$ - S) being integrated over some energy range; for the purpose of this work we chooze the range to be 10 to 18 keV. In this case J is a measure of distortion.

Note: in real measurement we obtain the S only; no information about S$_l$ is available there. One of possible technique is to fit S with S$_l$ in some energy range where the statistics is rather high; for the purpose of this work we chooze it to be 4-10 keV. The S is fitted with N (statistics) and R (resolution) being released.

Fig.4 - the model response of proportional counter to electrons of 5 and 15 keV. Primary electrons lines are shown in magenta; first it is spread with resolution R=16\% (at 5.9 keV), blue line. Next, the response is distorted by degraded events (due to wall and edge effects) - it looks like a left-handed shoulder (in blue) at each line with D=1\%; and finally it is distorted by pile-up with C=0.1\% (black line).

Fig.5 - the model tritium electron spectrum (blue) distorted by R=16\%, D=1\% and C=0.1\% (red) in propoprtional counter; during fit procedure D and C are usually fixed at the same central values D0=1\% and C0=0.1\%.

Systematic distortion is estimated further.

Fig.6 - the model difference (S$_l$ - S) in case of m$_s$=4 keV, U$^2$=10$^{-4}$ and N=10$^{12}$ is presented (circles with error); the S$_l$ indeed is (S$_l$)$_{fit}$ here, i.e. a result of fit of the S at 4-10 keV range. The distortion of model difference due to degraded events is also shown in color. Red line is the distortion due to fit with D=1.01xD0; blue one is due to D=0.99xD0.

Fig.7 - the same, but distorted due to pile up; red line is the distortion due to fit with C=1.01xC0; blue one is due to C=0.99xC0.

Fig.8 - the amplitude of systematic distortion J at zero mixing (U$^2$=0) is shown as a function of variaton of D (blue) and C (black) in the interval of $\pm$1\% around D0 and C0; the 3-sigma boundaries for N=10$^{12}$ are shown also.

The main conclusion is that we need the accuracy of D and C to be 0.2\% and 0.9\% resp. in order to keep the systematic uncertainty to be similar to statistic one. Such accuracy may be achieved in calibrating measurements with statistics as high as 10$^8$ events; it looks to have no problems.

\newpage

\end{document}